\documentclass[runningheads]{llncs}

\usepackage{amsmath}
\usepackage{mhchem}
\usepackage{graphicx}
\usepackage{braket}
\usepackage{dsfont}
\usepackage{xcolor}
\begin{document}
\title{Solving Quantum Chemistry Problems with a D-Wave Quantum Annealer }
%
%
\author{Michael Streif\inst{1} \and
Florian Neukart\inst{2,3} \and
Martin Leib\inst{1}}
\authorrunning{M. Streif et al.}
%
\institute{Data:Lab, Volkswagen Group, Ungererstr. 69, 80805 M{\"u}nchen, Germany\and
LIACS, Leiden University, Niels Bohrweg 1, 2333 CA Leiden, Netherlands\and
Volkswagen Group of America, 201 Post Street, San Francisco, CA 94108, United States of America}
\maketitle              
\begin{abstract}

Quantum annealing devices have been subject to various analyses in order to classify their usefulness for practical applications. While it has been successfully proven that such systems can in general be used for solving combinatorial optimization problems, they have not been used to solve chemistry applications. In this paper we apply a mapping, put forward by Xia et al. \cite{kais}, from a quantum chemistry Hamiltonian to an Ising spin glass formulation and find the ground state energy with a quantum annealer.  Additionally we investigate the scaling in terms of needed physical qubits on a quantum annealer with limited connectivity. To the best of our knowledge, this is the first experimental study of quantum chemistry problems on quantum annealing devices. We find that current quantum annealing technologies result in an exponential scaling for such inherently quantum problems and that new couplers are necessary to make quantum annealers attractive for quantum chemistry.

\end{abstract}

\section{Introduction}
Since the seminal talk by Richard Feynman \cite{feynman1982simulating} that jumpstarted the race for practical quantum computers, scientist have been dreaming of machines that can solve highly complicated quantum mechanical problems, which are inaccessible with classical resources. Molecules, such as caffeine, are already of such great complexity, that classical computers are incapable of simulating the full dynamics. Similar problems arise in the search for more efficient batteries, which is an important task for the upcoming electrification of traffic. Another prominent example is the simulation of photosynthesis processes in organic materials, which might lead to more efficient solar cells but is impossible to simulate due to the highly complex structure of the problem. The common ingredient which makes these problems so incredible hard to simulate is the exponentially growing Hilbert space. However a quantum computer, being a complicated quantum mechanical system itself, harnesses the power of the exponentially growing Hilbert space. Therefore, quantum computing is generally believed to be able to find solutions to these complex and important problems. An important class of such quantum problems are electronic structure problems \cite{lanyon2010towards,whitfield2011simulation}. There, the goal is to find the ground state energy of a quantum system consisting of electrons and stationary nuclei. To simulate these systems one has to solve the Schr{\"o}dinger equation, $H\ket{\Phi}=E \ket{\Phi}$, where $H$ is the Hamiltonian describing the dynamics of the electrons in presence of stationary cores. Solving the Schr{\"o}dinger equation can get unfeasible already when adding just another single electron to the system. Due to this exponentially increasing complexity of the problem solving this type of problem belongs to the most difficult calculations in both science and industry. State-of-the-art classical computers are able to solve such problems exactly for smaller molecules only. Various numerical methods such as Hartree-Fock (HF), quantum Monte Carlo (qMC), density functional theory (DFT) or configuration interaction methods (CI) were developed to give an approximate solution to these kinds of problems. In the future, quantum computers will be able to find exact solutions to these problems and thereby obtaining a deeper insight into nature.

In recent years early industrial  incarnations of gate based quantum computers are appearing \cite{barends2016digitized,dicarlo2009demonstration,debnath2016demonstration,reagor2018demonstration} and small quantum chemistry algorithms have already been executed on these machines, such as the Variational Quantum Eigensolver (VQE) \cite{peruzzo2014variational} or the Phase Estimation Algorithm (PEA) \cite{kitaev1995quantum}, which utilize the possibility to represent electrons in atomic orbitals as qubits on the quantum processor. Kandala et al. used transmon qubits to simulate molecular hydrogen, lithium hydride and beryllium oxide \cite{kandala2017hardware}. Hempel et al. used a trapped ion quantum computer to simulate molecular hydrogen and lithium hydride \cite{hempel2018quantum}. However, current gate model devices suffer from various shortcomings, e.g. a small number of qubits, errors caused by imperfect gates and qubits, and decoherence effects. These effects limit the coherence time and consequently the total number of gates which can be applied before decoherence effects destroy the potential for any quantum speedup. Research is therefore focusing on finding hardware-efficient, i.e. shallow circuits for solving electronic structure problems \cite{kandala2017hardware,kivlichan2018quantum,babbush2018quantum,babbush2017low}. 

Parallel to the development of gate based quantum computers there have been efforts to build quantum annealers on an industrial scale since the turn of the century \cite{johnson2011quantum} and the number of qubits in such devices have rapidly increased over the last few years. Quantum annealers use a quantum enhanced heuristic to find solutions to combinatorial optimization problems, such as traffic flow optimization \cite{neukart2017traffic} or cluster analysis \cite{neukart2018quantum}. In more recent time, quantum annealing systems were utilized to sample from a classical or quantum Boltzmann distribution, which enabled using such machines for machine learning purposes such as the training of deep neural networks \cite{adachi2015application} or reinforcement learning \cite{levit2017free}. Despite their higher maturity with respect to gate based quantum computers, quantum annealers have, to the best of our knowledge not been used to solve quantum chemistry problems yet. In this present contribution, we follow an approach put forward by Xia et al. \cite{kais} to map electronic structure Hamiltonians to a classical spin glass and subsequently find the ground state with quantum annealing.

We start in Sec.~\ref{QANS} with a short introduction into quantum annealing followed by  a general presentation of the electronic structure problem and a recapitulation of the approach proposed by \cite{kais} in Sec.~\ref{electronicstructure}. In Sec.~\ref{sec:level2}, we present first results obtained with the D-Wave 2000Q machine for molecular hydrogen (\ce{H2}) and lithium hydride (\ce{LiH}). In Sec.~\ref{methods}, we give technical details of the obtained results and, finally, in Sec.~\ref{sec:discussion}, we conclude with a summary of our findings and an outlook of possible future research directions. 
\section{Quantum Annealing in a nutshell}
\label{QANS}
Quantum annealing belongs to a class of meta-heuristic algorithms suitable for solving combinatorial optimization problems. The quantum processing unit (QPU) is designed to find the lowest energy state of a spin glass system, described by an Ising Hamiltonian,
\begin{align}
	H_{\mathrm{SG}}=\sum_{i}h_i\sigma_z^i+\sum_{i,j}J_{ij}\sigma_z^i\sigma_z^j,
	\label{ising}
\end{align}
where $h_i$ is the on-site energy of qubit $i$, $J_{ij}$ are the interaction energies of two qubits $i$ and $j$, and $\sigma_z^i$ is the Pauli matrix acting on the $i$-th qubit. Finding the ground state of such a spin glass system, i.e. the state with lowest energy, is a NP problem. Thus, by mapping other NP problems onto spin glass systems, quantum annealing is able to find the solution of them. The idea of quantum annealing is to prepare the system in the ground state of a Hamiltonian which is known and easy to prepare, e.g. $H_\mathrm{X}=\sum_i \sigma_x^i$. Then we change the Hamiltonian slowly such that it is the spin glass Hamiltonian at time T,
\begin{align}
    H(t)=\left(1-\frac{t}{T}\right)H_X+\left(\frac{t}{T}\right)H_{SG}.
\end{align}
If $T$ is long enough, according to the adiabatic theorem, the system will be in the ground state of the spin glass Hamiltonian $H_{\mathrm{SG}}$. 
\section{The electronic structure problem and its mapping on an Ising Hamiltonian}
\label{electronicstructure}
\subsection{The electronic structure problem}
The behaviour of electrons inside molecules is determined by their mutual interaction and the interaction with the positively charged nuclei. To describe the dynamics of the electrons or to find their optimal energetic configuration one has to solve the Schr{\"o}dinger equation of this many-body quantum system. The Hamiltonian for a system consisting of $M$ nuclei and $N$ electrons can be written in first quantization as 
\begin{align}
\label{equation2}
H=&-\sum_i \frac{\nabla_i^2}{2} -\sum_A\frac{\nabla_A^2}{2M_A}-\sum_{i,A}\frac{Z_A}{\left|r_i-R_A\right|}+\sum_{i,j}\frac{1}{\left|r_i-r_j\right|}+\sum_{A,B}^M\frac{Z_AZ_B}{\left|R_A-R_B\right|}  ,
\end{align}
where $r_i$ are the positions of the electrons, $M_A$, $R_A$ and $Z_A$ are the mass, position and  charge in atomic units of the nuclei respectively. Using the second quantization, i.e. writing Eq.~\ref{equation2} in terms of fermionic creation and annihilation operators $a_i^\dagger$ and $a_j$ of a specific fermionic mode $i$ and $j$ with $\{a_i,a_j^\dagger\}=\delta_{ij}$ and applying  the Born-Oppenheimer approximation, which assumes that the nuclei do not move due to their much greater mass ($M_A\gg1$), we get the following Hamiltonian
\begin{align} 
H=\sum_{i,j}h_{ij}a_j^{\dagger}a_i+\frac{1}{2}\sum_{i,j,k,l}h_{ijkl}a_i^{\dagger}a_j^{\dagger}a_ka_l.
\label{equation6}
\end{align}
The parameters $h_{ij}$ and $h_{ijkl}$ are the one- and two-particle integrals for a specific, problem-dependent basis set $\ket{\psi_i}$, which has to be appropriately chosen,
\begin{align}
h_{ij}&=\braket{\psi_i|\left(-\frac{\nabla_i^2}{2}-\sum_{A}\frac{Z_A}{\left|r_i-R_A\right|}\right)|\psi_j},\nonumber\\
h_{ijkl}&=\braket{\psi_i\psi_j|\frac{1}{\left|r_i-r_j\right|}|\psi_k\psi_l}.
\label{onetwoparticleintegrals}
\end{align}
Quantum devices utilize qubits, i.e. we have to find a qubit representation of the fermionic Hamiltonian. The Jordan-Wigner or Bravyi-Kitaev transformation \cite{seeley2012bravyi} are the most prominent examples achieving this task and map the fermionic creation and annihilation operators to Pauli matrices, both leading to a qubit Hamiltonian of the form
\begin{align}
\label{equation7}
H=&\sum_{i,\alpha}h_{\alpha}^i\sigma_{\alpha}^i+\sum_{i,j,\alpha,\beta}h_{\alpha\beta}^{ij}\sigma_{\alpha}^{i}\sigma_{\beta}^j+\sum_{i,j,k,\alpha,\beta,\gamma}h_{\alpha\beta\gamma}^{ijk}\sigma_{\alpha}^i\sigma_{\beta}^j\sigma_{\gamma}^k+\dots
\end{align}
where $\sigma^i_{\alpha=x,y,z}$ are the Pauli matrices acting on a single qubit at site $i$. This qubit Hamiltonian now encodes the electronic structure problem we would like to solve by using a quantum annealer. 
\subsection{\label{sec:level1}Formulation as an Ising spin glass}
\label{implementation}

To find the ground state energy of an electronic structure problem with a quantum annealing device, it is necessary to find a representation of the problem, cf. Eq.~\ref{equation7}, in the form of a classical spin glass system described by the Ising Hamiltonian, cf. Eq.~\ref{ising}. To accomplish this, we use the method proposed by Xia et al. \cite{kais}. In this section, we shortly recapitulate the method, for a detailed overview, cf.~\cite{kais}.

Terms containing $\sigma_x$ and $\sigma_y$ in the electronic structure Hamiltonian prohibit a direct embedding on a quantum annealer. To overcome this problem, the idea is to introduce $r$ ancillary qubits for all $n$ qubits of the original Hamiltonian respectively. Each Pauli operator is mapped to a diagonal operator in the classical subspace of the larger $(r\times n)$-qubit Hilbert space using the following relations:
\begin{align}
&\sigma_x^i\rightarrow\frac{1-\sigma_z^{i_j}\sigma_z^{i_k}}{2}S(j)S(k)&\sigma_y^i&\rightarrow \mathrm{i}\frac{\sigma_z^{i_k}-\sigma_z^{i_j}}{2}S(j)S(k)\nonumber\\
&\sigma_z^i\rightarrow \frac{\sigma_z^{i_j}+\sigma_z^{i_k}}{2}S(j)S(k)&\mathds{1}&\rightarrow\frac{1+\sigma_z^{i_j}\sigma_z^{i_k}}{2}S(j)S(k)
\end{align}
In these mappings, $\sigma_{\alpha=x,y,z}^{i_j}$ denote the Pauli matrices acting on the j-th ancillary qubit of the i-th original qubit. The mapped operators then reproduce the action of the former operators on the wavefunction in the original Hilbert space. The function $S(j)$ takes care of the sign of the state in the original Hilbert space. 
By increasing $r$, we are able to grasp more quantum effects and therefore are able to get a more precise estimation of the true ground state energy. In the following, we use this mapping and the proposed algorithm in \cite{kais} to find a classical spin glass approximation of our electronic structure Hamiltonian, which we then solve by using the methods of quantum annealing.

\section{\label{sec:level2}Results from the D-Wave 2000Q}
In this section, we present first examples of electronic structure calculations done on a physical quantum annealing device, namely the D-Wave 2000Q machine. This quantum annealing device has 2048 qubits and 6016 couplers, arranged in a Chimera-type structure.  Further information about the precise architecture and technical details can be found in \cite{dwave2000q}. 

We calculate the ground state energies of molecular hydrogen (\ce{H2}) and lithium hydride (\ce{LiH}) for various interatomic distances. Moreover, we calculate the number of required qubits when using a quantum annealing device with limited connectivity. For an overview of our used methods, we refer to Sec.~\ref{methods}.
\subsection{Molecular hydrogen - \ce{H2}}
\begin{figure}
	\centering
	\makebox[\textwidth][c]{\includegraphics[width=1\columnwidth]{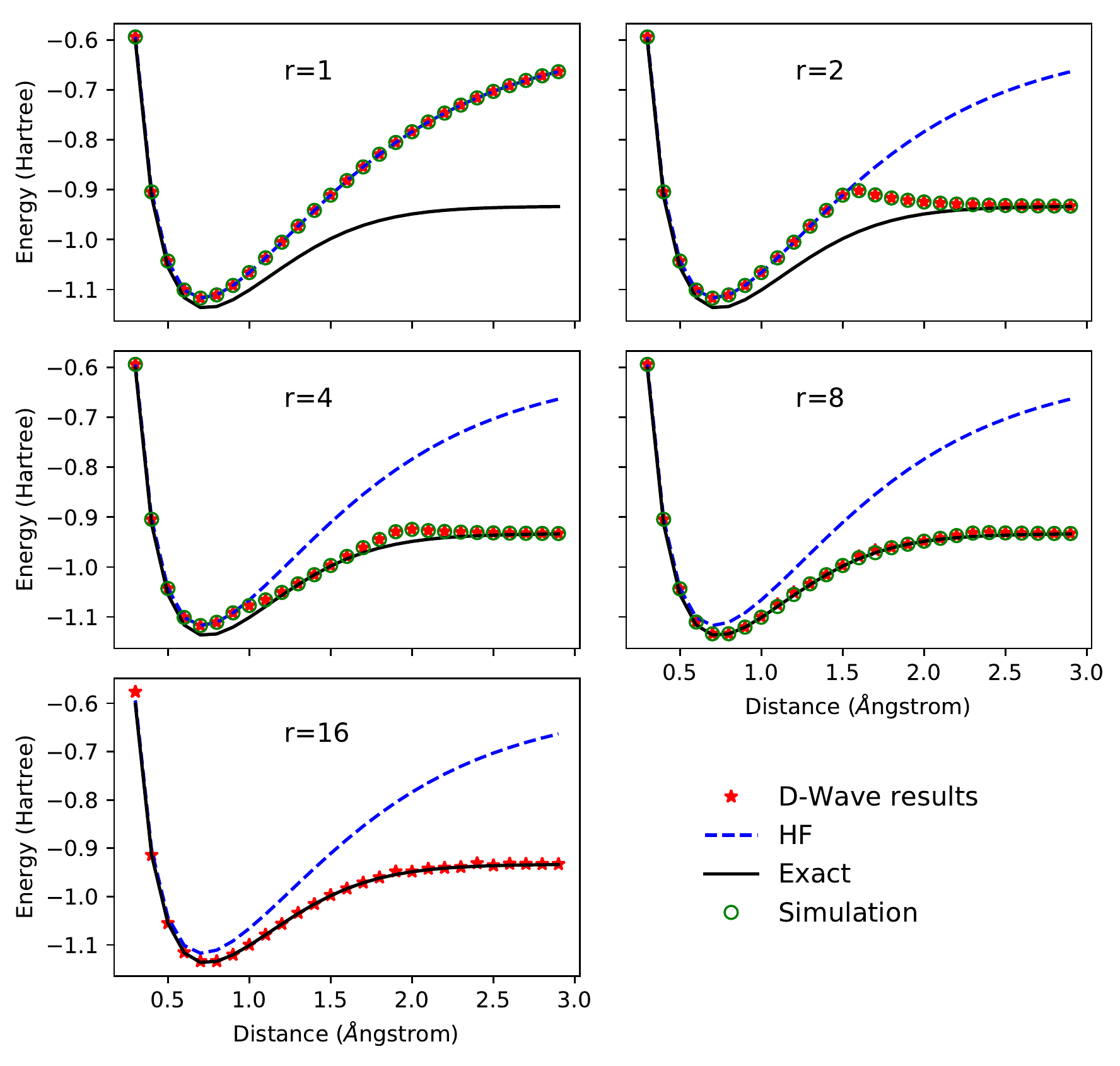}}
	\caption[Simulation of molecular hydrogen]{Ground state energies of molecular hydrogen $\ce{H2}$ for various interatomic distances and different values of the scaling factor $r$. The red asterisks show the results from D-Wave 2000Q, the blue line shows the Hartree-Fock energy of the Hamiltonian, the black line the  ground state energy obtained by the exact diagonalization of the molecular Hamiltonian, and the green circles show the simulated results of the classical $(r\times n)$-qubit Hamiltonian. As expected, by increasing the value of $r$, we increase the accuracy of the results. The D-Wave quantum annealing device closely reproduces the simulated results, meaning that it was able to find the right ground state energy for the given problem. For $r=16$ we are very close to the exact results, where in this case we were not able to do the numerical calculation, thus we show  experimental results only. In Fig.~\ref{fig:scalingH2} we show the required qubits on the quantum processor for each of these  plots. For all the experiments presented in this plot, we used an annealing time of $\tau=100\mu s$ and $1000$ annealing runs.}
	\label{fig:h2dwave}
\end{figure}

\begin{figure}
	\centering
	\includegraphics[width=0.8\columnwidth]{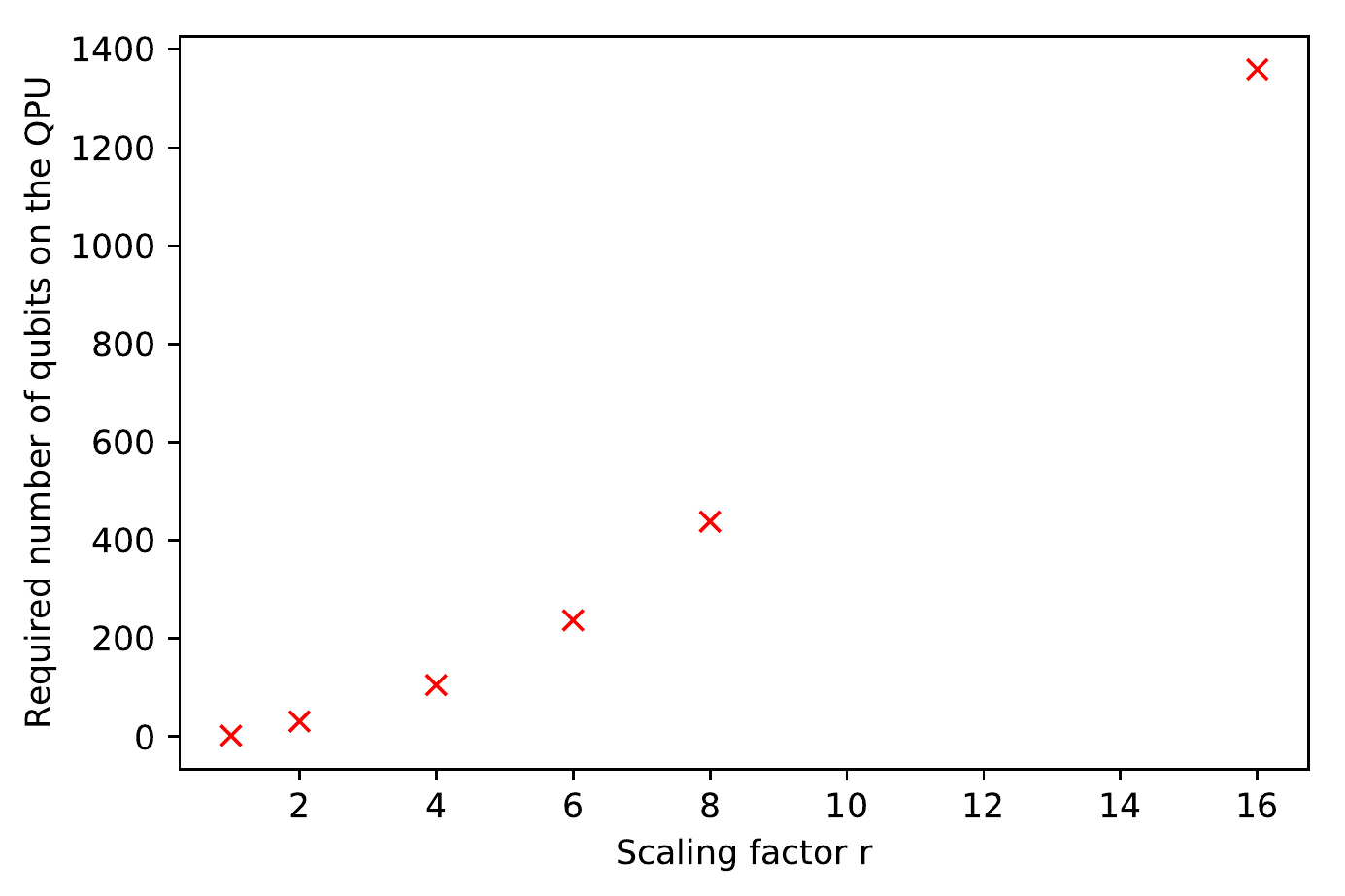}
	\caption[Scaling of molecular hydrogen ]{Here we show the number of required qubits after embedding the classical $(r\times n)$-qubit Hamiltonian of molecular hydrogen on the real device for increasing scaling factors $r$.}
	\label{fig:scalingH2}
\end{figure}
We start with molecular hydrogen, $\ce{H2}$, which has already been the testbed for first electronic structure calculations on gate model devices \cite{kandala2017hardware,hempel2018quantum}. With the two electrons of molecular hydrogen we have to account for 4 atomic orbitals. After calculating the fermionic Hamiltonian, a Bravyi-Kitaev transformation yields the qubit Hamiltonian,
\begin{align}
H_{\ce{H2}}=&g_1 \mathds{1} +
g_2 \sigma_z^0 \sigma_z^2 \sigma_z^3 +
g_3 \sigma_z^0 \sigma_z^2 +
g_4 \sigma_z^2 +
g_5 \sigma_y^0 \sigma_z^1 \sigma_y^2 \sigma_z^3
+g_6 \sigma_z^0 \sigma_z^1 \sigma_z^2
\nonumber
+g_7 \sigma_z^0 \sigma_z^1 \\
&+g_8 \sigma_z^1 \sigma_z^3 +
g_8 \sigma_z^1 \sigma_z^2 \sigma_z^3
+g_{10} \sigma_z^0 \sigma_z^1 \sigma_z^2 \sigma_z^3 +
g_{11} \sigma_z^1 +
g_{12} \sigma_z^0 +
\nonumber
g_{13} \sigma_x^0 \sigma_z^1 \sigma_x^2\\ 
&+g_{14} \sigma_x^0 \sigma_z^1 \sigma_x^2 \sigma_z^3 +
g_{15} \sigma_y^0 \sigma_z^1 \sigma_y^2 .
\label{H2Ham}
\end{align}
In this expression, $g_i$ are parameters calculated from the one- and two-particle integrals, i.e. Eq.~\ref{onetwoparticleintegrals}. These parameters depend on the interatomic distance $R$ between both hydrogen atoms, $g_i=g_i(R)$. As shown in \cite{o2016scalable}, the first and third qubit in this Hamiltonian does not affect the population numbers of the Hartree-Fock state $\ket{\Psi_{\mathrm{HF}}}$. Therefore both qubits are not important for finding the ground state energy and can be neglected in the remainder of this calculation, yielding the 2-qubit Hamiltonian
\begin{align}
H_{\ce{H2}}=&g_0 \mathds{1} +
g_1 \sigma_z^0 +
g_2 \sigma_z^1 +
g_3 \sigma_z^0 \sigma_z^1 +
g_4 \sigma_x^0 \sigma_x^1 +
g_4 \sigma_y^0 \sigma_y^1
\end{align}
Exact values for $g_i$ can be found in \cite{o2016scalable}. To find the the ground state of this Hamiltonian, we map it on a  $(r\times n)$-qubit Hamiltonian and find a classical 4-local spin glass representation in this larger Hilbert space. We then reduce the 4-local to 2-local terms by again introducing ancillary qubits and subsequently find the lowest eigenvalue of this 2-local classical spin glass Hamiltonian by embedding it on the Chimera graph of the D-Wave 2000Q machine. We repeat these experiments for different values of the interatomic distance and for different scaling factors $r$.

The results of these experiments are shown in Fig.~\ref{fig:h2dwave}. For $r=1$, we merely remove any term in the Hamiltonian that contains $\sigma_x$ or $\sigma_y$ Pauli operators. As we start with a Hamiltonian which converged after a classical Hartree-Fock calculations, see Sec.~\ref{methods}, this case should provides us with Hartree-Fock energies. For $r=2$, we have to account for 20 terms in total and 31 qubits on the QPU after embedding. For the largest scaling factor we used, $r=16$, we have to use 1359 qubits on the QPU to account for 1490 terms in the 2-local spin glass version of the Hamiltonian. 

In Fig.~\ref{fig:scalingH2}, we show the number of used qubits on the quantum annealing processor for different $r$. These numbers are the result of the mapping to a Hamiltonian containing only $\sigma_z$ operators, the mapping to two-local terms and the embedding on the chimera structure of the D-Wave 2000Q device. The required number of qubits approximately scales quadratically with the scaling factor. We note that this does not imply that the whole method scales quadratically when going to larger systems, as we might need an exponential increasing scaling factor for reaching an sufficient accuracy.  
\subsection{Lithium hydride - \ce{LiH}}
\begin{figure}
	\centering
	\makebox[\textwidth][c]{\includegraphics[width=1\columnwidth]{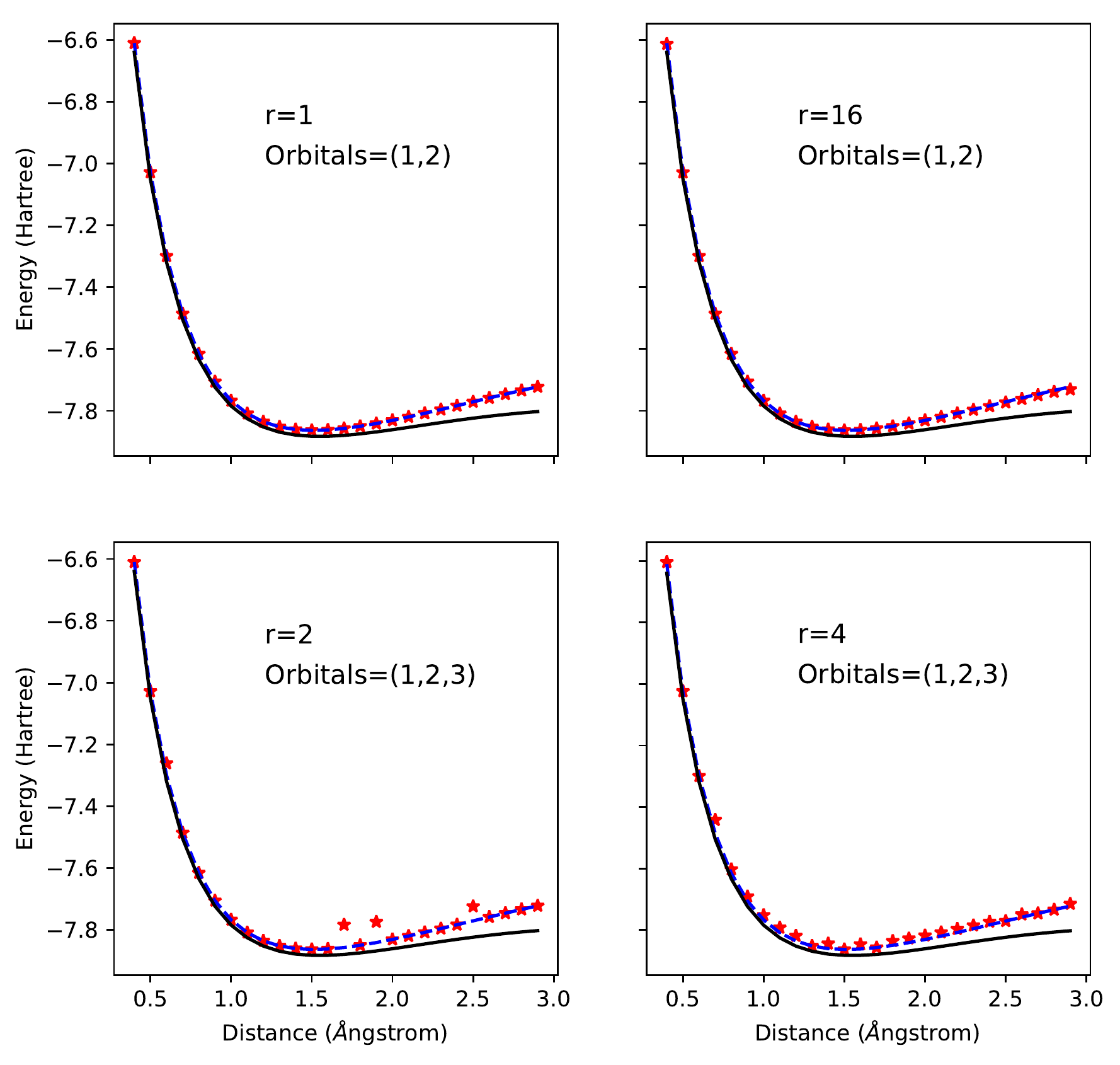}}
	\caption[Simulation of lithium hydride]{Ground state energies of lithium hydride, $\ce{LiH}$, for different active orbitals, different values of $r$ and various interatomic distances. $\ce{LiH}$ has 6 orbitals, here numbered from $0$ to $5$, plus spin degrees of freedom. The red asterisks show the results from D-Wave 2000Q, the blue line shows the Hartree-Fock energy of the Hamiltonian and the black line the exact ground state energy. For the sake of brevity, we do not show numerical results of the transformed Hamiltonian here. For the experiments with 2 orbitals, we used an annealing time of $\tau=100\mu s$ and $1000$ annealing runs, whereas for 3 orbitals, we used an annealing time of $\tau=100\mu s$ and $9000$ annealing runs.}
	\label{fig:lihdwave}
\end{figure}
To gauge the potential of the presented technique we start considering more complicated molecules such as lithium hydride. Lithium hydride has four electrons and we have to account for 12 orbitals in the minimal basis. We follow the same steps as for molecular hydrogen, i.e. we derive the qubit Hamiltonian of the problem using a Bravyi-Kitaev transformation of the fermionic Hamiltonian. We again are able to truncate the Hamiltonian to a smaller space by exploiting the fact that two qubits do not change its population numbers when starting in a Hartree-Fock state, leaving us with a 10-qubit Hamiltonian. Using the transformation given in Sec.~\ref{implementation}, we map the Hamiltonian to a classical spin glass. 
In contrast to molecular hydrogen, for lithium hydride, due to limited number of qubits of the quantum hardware and the embedding overhead due to the limited connectivity, we are not able to take all atomic orbitals into account. Therefore we use an active space representation, i.e. we freeze orbitals and optimize the electron configuration in the remaining orbitals. After finding a 2-local representation of the remaining Hamiltonian and an embedding on the Chimera graph, we again find the ground state by the use of quantum annealing. For the sake of brevity, we do not state the full Hamiltonian here. In the following we present results for both various atomic orbitals and scaling factors.

These results can be found in Fig.~\ref{fig:lihdwave}. The results of all shown plots are very close to the initial Hartree-Fock energies, meaning that we were not able to improve from the initial Hartree-Fock energies, which was the starting point of our calculation. Additionally, as the problem gets more complicated, D-Wave was in some cases not able to find the true ground state energy of the transformed Hamiltonian. For a scaling factor $r=4$ and 3 orbitals, we have to use 1558 qubits on the QPU. In our experiments, it was not possible to use more orbitals while $r\neq 1$.
\section{Methods}
\label{methods}
In this section, we shortly summarize the technical details of our experiments. For both molecules, $\ce{H2}$ and $\ce{LiH}$, we start by determining the fermionic Hamiltonian by calculating the one- and two-particle integrals, cf. Eq.~\ref{onetwoparticleintegrals} using the Psi4 module of Google's OpenFermion library \cite{mcclean2017openfermion}. We used the molecular wavefunctions from converged Hartree-Fock calculations obtained by using a minimal basis set, namely STO-3G, which are created from 3 Gaussian orbitals fitted to a single Slater-type orbital. We then apply a Bravyi-Kitaev transformation to map the second-quantized fermionic operators onto Pauli matrices to obtain a qubit representation of the problem. By using the method described in Sec.~\ref{implementation}, we map the n-qubit Hamiltonian to a $(r\times n)$-qubit Hamiltonian. As D-Wave's implementation consists of 2-local terms only, we introduce ancillary qubits to find a 2-local representation of the Hamiltonian to embed the problem subsequently  onto the Chimera graph structure of the quantum annealing device. To make sure that the embedding is optimal, we use the heuristic algorithm provided by D-Wave and generate 100 random embeddings for each bond length and find the ground state energy for each of these 100 embeddings with the D-Wave 2000Q. We then only keep the best solution, i.e. the solution with lowest energy. To compare our results with classical methods, we use the converged Hartree-Fock energies and the exact results, which we obtained by a numerical diagonalization of the qubit Hamiltonian.
\section{\label{sec:discussion}Conclusion \& Outlook}             

In this paper, we did a first examination of quantum chemistry problems on the D-Wave 2000Q by using an approach proposed by Xia et al. \cite{kais}. We were able to calculate the ground state energies of molecular hydrogen and lithium hydride with the current generation of the QPU. For molecular hydrogen, $\ce{H2}$, our ground state energy estimations were very close to the exact energies when going to scaling factor of $r=16$. For achieving this accuracy, we already had to use a large fraction of available qubits on the quantum processor. We moreover showed a first scaling of this methods under real conditions and overheads of quantum annealing hardware. For lithium hydride, \ce{LiH}, we were not able to reproduce closely the ground state energy with the currently available hardware. When accounting for $3$ orbitals and using a scaling factor of $r=4$, we already had to use 1558 qubits, which is a large fraction of available qubits. To summarize: the investigated method in general works, but it might be difficult to apply it to larger systems.

However, we give some further research ideas how quantum annealing devices could be applied to quantum chemistry problems in the nearer future. Quantum annealing devices which utilize interactions beyond the standard Ising Hamiltonian, i.e. beyond $\sigma_z\otimes\sigma_z$ interactions, could be helpful as they would allow to use a larger fraction of the Hilbert space. When having access to the right interactions, an efficient embedding onto quantum annealing processors could be feasible \cite{babbush2014adiabatic}. Another possibility is to use the recently announced new features of the D-Wave machine, such as the possibility to do reverse annealing or to stop the annealing process in an intermediate point of the adiabatic evolution, i.e. between start and final Hamiltonian. This may be utilized for getting access to terms which are non-diagonal in the computational basis and in the end could enable to sample from low-lying energy states. Together with a classical subroutine, one could find the solution of the problem. Another possibility could be to use the D-Wave machine to calculate Hartree-Fock, i.e. approximate energies of the problem. Together with a classical loop, quantum annealing devices could be used to estimate the Hartree-Fock energy of large molecules, which would be unfeasible with classical resources. Another promising alternative is to use machine learning to improve the found ground state energies.

\section*{\label{sec:level6}Acknowledgments}
We thank VW	Group	CIO	Martin	Hofmann	and	VW	Group	Region	Americas	CIO	Abdallah Shanti, who	 enable	 our	 research.	Any opinions, findings, and conclusions expressed in this paper do not necessarily reflect the views of the Volkswagen Group.
%
%
%
\bibliographystyle{splncs04}
\bibliography{mybibliography}
\end{document}